\documentclass[%
  reprint,
  superscriptaddress,
  amsmath,amssymb,
  aps,
  pre,
  floatfix,
]{revtex4-2}

\usepackage[T1]{fontenc}
\usepackage{graphicx}
\graphicspath{{figs/}}
\usepackage{dcolumn}
\usepackage{bm}
\usepackage{silence}
\usepackage{hyperref}
\makeatletter
\def\@bibdataout@aps{%
  \immediate\write\@bibdataout{%
    @CONTROL{%
      apsrev42Control%
      \longbibliography@sw{%
        ,author="48",editor="1",pages="0",title="0",year="1"%
      }{%
        ,author="48",editor="1",pages="0",title="",year="1"%
      }%
    }%
  }%
  \if@filesw
  \immediate\write\@auxout{\string\citation{apsrev42Control}}%
  \fi
}%
\makeatother

\begin{document}

\title{
  Portfolio Allocation under Heterogeneous Scales and Multifractality
}

\author{Shinji Kakinaka}%
\email{kakinaka.shinji@kochi-tech.ac.jp}
\affiliation{%
  School of Economics and Management, Kochi University of Technology, Japan
}%
\author{Ken Umeno}%
\email{umeno.ken.8z@kyoto-u.ac.jp}
\affiliation{%
  Course of Applied Mathematics and Physics, Graduate School of Informatics, Kyoto University, Japan
}%

\date{November 25, 2026}

\begin{abstract}
  Cross-correlations between financial signals are neither scale-free nor amplitude-independent: they vary with the time scale over which they are measured and with the magnitude of the fluctuations that dominate the average. We exploit this structure to construct a portfolio allocation model in which the risk functional is the signed fluctuation function of multifractal cross-correlation analysis (MFCCA), indexed by a scale $s$ and a fluctuation order $q$. Unlike MFDCCA-type criteria, which rectify local detrended covariances before aggregation, MFCCA retains their sign, so that co-moving and counter-moving components contribute to risk with opposite signs; for $q=2$ the resulting quadratic form coincides with the detrended fluctuation function of the portfolio series itself, recovering the mean--variance criterion as a scale-dependent limit.
  Using two-component ARFIMA and Markov-switching multifractal processes, we show that prescribed multiscale and multifractal dependence is transmitted into the optimal weights, and that sign preservation contributes more to the reduction of tail risk than aggregation over fluctuation orders. Applied to financial multi-assets, the criterion lowers drawdown, Value-at-Risk, and expected shortfall relative to the mean--variance benchmark at every required return, in and out of sample, without any loss in realized portfolio return.
  The construction maps signed multiscale interaction structures onto resource-allocation decisions,
  which is applicable to complex financial systems whose components interact across heterogeneous scales with amplitude-dependent coupling.
\end{abstract}

\maketitle

\section{\label{sec:introduction}Introduction}

Modern Portfolio Theory (MPT) provides the standard mathematical framework for constructing diversified portfolios by balancing expected return against risk~\citep{Markowitz1952}. Its mean--variance formulation is attractive because it is transparent and analytically tractable, but the variance--covariance matrix summarizes dependence only through second moments and linear co-movements. In empirical applications, optimal allocations can therefore be sensitive to estimation error, unstable covariance structures, and distributional features of asset returns such as heavy tails, volatility clustering, and nonlinear dependence~\citep{Michaud1989, Mandelbrot1963, Cont2001}. These limitations do not invalidate the risk--return principle of MPT; rather, they motivate portfolio criteria that preserve this principle while replacing the conventional covariance measure with a richer dependence measure.

The Fractal Market Hypothesis (FMH) provides one such perspective by emphasizing that financial markets are composed of heterogeneous investors operating over different investment horizons~\citep{Peters1994}. Under this view, market stability and liquidity depend on the coexistence of agents with different time-scale preferences, and changes in their relative dominance can produce scale-dependent risk and dependence structures. Financial returns may also exhibit multifractal behavior, meaning that small and large fluctuations can obey different scaling laws~\citep{CalvetFisher2002}. Consequently, portfolio diversification should be examined not only across assets, but also across time scales and fluctuation magnitudes.

Existing fractal-based portfolio studies can be broadly classified into several streams. First, multiscale CAPM and fractal-regression approaches incorporate heterogeneous investment horizons into asset pricing and efficient-frontier construction~\citep{Kristoufek2018, Kristoufek2018_CAPM, Tilfani2019, Tilfani2020}. Second, detrended cross-correlation analysis (DCCA)-based portfolio models replace the covariance or correlation matrix in the mean--variance framework with scale-dependent detrended cross-correlation measures~\citep{Sun2016DCCAportfolio, CHUN2020, Zhang2022, Kakinaka2023}. Third, multifractal extensions, especially mean-MFDCCA portfolio models, incorporate both time scales and fluctuation orders into portfolio construction by using multifractal detrended cross-correlation functions~\citep{Li2021MeanMFDCCA, Li2024MultiAssetMD}. Related studies have also proposed fractal statistical measures under power-law return distributions and alternative multifractal cross-correlation frameworks, including MF-X-DMA-based portfolio selection and multiwavelet denoising integrated methods~\citep{Wu2021FractalStatistical, Wang2021MeanMFXDMA, Jiang2011,  Zhu2020MultiwaveletMFDCCA}. Recent empirical studies using related MFDCCA-type tools further document multifractal cross-correlations in stock, commodity, green-bond, and emerging-market financial systems~\citep{Chen2024ChinaUSMFDCCA, Acikgoz2024GreenBondCommodityMFDCCA, Acikgoz2025EmergingFinancialMFDCCA}.

Multifractal cross-correlation analysis (MFCCA) is particularly useful for this purpose because it generalizes detrended cross-correlation analysis to arbitrary fluctuation orders while preserving the sign of local detrended covariance~\citep{Podobnik2008, Owiecimka2014}. In this framework, the scale parameter $s$ represents the time horizon over which dependence is measured, whereas the fluctuation order $q$ controls whether the analysis emphasizes relatively small or large fluctuations. This makes it possible to examine whether cross-asset dependence changes across both investment horizons and fluctuation magnitudes.
The amplitude dependence is also visible at the level of a whole market. When the correlation matrix of stock or cryptocurrency returns is computed with the $q$-dependent detrended cross-correlation coefficient, the minimum spanning tree obtained for medium-sized fluctuations ($q=2$) differs in topology from the one obtained for large fluctuations ($q=4$), and the two evolve differently over time scales~\cite{Kwapien2017qMST, Watorek2025}. W{\k{a}}torek et al.~\cite{Watorek2025} argue that allocation and risk management should take such amplitude-specific structures into account. Unlike MFDCCA-type procedures that transform local fluctuation components into non-negative quantities before aggregation~\citep{Zhou2008}, MFCCA retains directional co-movement information through a signed fluctuation function.
This distinction is important for portfolio allocation. Diversification depends not only on the strength of co-movement but also on its direction: positive co-movement tends to increase joint risk, whereas weak or negative co-movement can reduce portfolio risk. If the sign of local detrended covariance is removed or obscured, economically relevant dependence structures may be masked even when multifractal scaling is detected. This motivates the use of MFCCA as a portfolio risk function. By preserving signed local covariance components while allowing both scale and fluctuation-order dependence, the proposed mean-MFCCA model is designed to reflect positive and negative co-movement more directly than mean-MFDCCA-type models.

This study contributes to the literature in four ways. First, we construct a mean-MFCCA portfolio model that embeds the sign-preserving MFCCA fluctuation function into the mean--variance portfolio framework. The key distinction from mean-DCCA and mean-MFDCCA benchmarks is that the proposed risk criterion preserves the sign of local detrended covariance, allowing the direction of asset co-movement to enter portfolio risk evaluation. Second, we formulate the model for multiple risky assets and aggregate optimal allocations across both time scales and fluctuation orders, so that heterogeneous investment horizons and heterogeneous fluctuation amplitudes are incorporated simultaneously. Third, we compare the model with mean-DCCA, mean-MFDCCA, and conventional mean--variance portfolios using synthetic time series designed to isolate multiscale and multifractal diversification effects, which allows the contributions of sign preservation and fluctuation-order aggregation to be evaluated separately. Fourth, we evaluate the empirical usefulness of the proposed model using practical risk measures, including drawdown, Value-at-Risk (VaR), and Expected Shortfall (ES), in both in-sample and out-of-sample settings.
From a complex-systems perspective, the proposed model provides a mapping from multiscale interaction patterns among financial signals to allocation outcomes.

The remainder of this paper is organized as follows. Section~\ref{sec:methodology} introduces the MFCCA and MFDCCA procedures and formulates the mean-MFCCA portfolio model. The following section investigates the diversification performance of the proposed model using synthetic time series with controlled multiscale and multifractal properties. We then apply the framework to empirical financial data and evaluate both in-sample and out-of-sample portfolio performance. Finally, the concluding section summarizes the main findings and discusses possible extensions.

\section{\label{sec:methodology}Methodology}
In this section, we present the multifractal cross-correlation analysis, and then develop a portfolio allocation model that optimizes the trade-off between expected return and a multifractal risk measure.

\subsection{\label{sec:MFCCA}Multifractal cross-correlation analysis}

MFCCA is an effective method for detecting the multifractal cross-correlation characteristics between two different simultaneously recorded time series, as well as the multifractal characteristics of a single time series~\cite{Owiecimka2014}.
The method is developed from the original procedure of the DCCA~\cite{Podobnik2008}, and therefore the initial steps are basically the same as the DCCA algorithm.

For a pair of given time series $\{x_t\}_{t=1}^N$ and $\{y_t\}_{t=1}^N$, we first construct the profiles $X(t)=\sum_{i=1}^t (x_i - \bar{x})$ and $Y(t)=\sum_{i=1}^t (y_i - \bar{y})$, where the bar denotes the average over the entire series. The profiles are divided into $N_s = \lfloor N/s \rfloor$ non-overlapping segments of length $s$. Since $N/s$ is not always an integer, the division is repeated from the other end to make sure no data is omitted from the procedure, so we have $2N_s$ segments in total for each series.
Next, the trend-eliminated series are obtained by applying a degree-2 polynomial least-squares fit within local segments $v=1,\ldots,2N_s$, and the local detrended cross-covariance is calculated as
\begin{align}
  \label{eq:fxy}
  f_{XY}^2(s, v) = \frac{1}{s} \sum_{t=1}^{s}\left\{X_v(t)-\tilde{X}_v(t)\right\}\left\{Y_v(t)-\tilde{Y}_v(t)\right\},
\end{align}
where $\tilde{X}_v(t)$ and $\tilde{Y}_v(t)$ denote the local trends of $v$th segment.
Finally, by taking the generalized average over all segments, the $q$th-order fluctuation function $F_{XY}^q(s)$ is calculated
\footnote{Following the notation of O{\'s}wie{\c{c}}imka et al.~\cite{Owiecimka2014}, we denote this function by $F^q_{XY}(s)$.}.
A direct generalization of this function is
\begin{align}
  \label{eq:Fxy}
  F^q_{XY}(s) &= \frac{1}{2N_s} \sum_{v=1}^{2N_s}\left \{ f_{XY}^2(s, v) \right \}^{q/2},
\end{align}
however, this representation has crucial issues towards practical use. The problem is that $f_{XY}^2(s, v)$ can take negative values, and hence $F^q_{XY}(s)$ is defined only when $q$ is an even integer value (otherwise an imaginary space should be introduced).
To deal with this, the method of MFCCA takes the sign of $f_{XY}^2(s, v)$ into account, allowing for any hierarchical order in the function~\cite{Owiecimka2014}. The $q$th-order fluctuation function is upgraded as
\begin{align}
  \label{eq:Fqxy_MFCCA}
  F^q_{XY,\mathrm{MFCCA}}(s) &= \frac{1}{2N_s} \sum_{v=1}^{2N_s}\mathrm{sgn}\left [ f_{XY}^2(s, v)\right ] \left | f_{XY}^2(s, v) \right |^{q/2},
\end{align}
for any $q\neq0$, and
\begin{align}
  \label{eq:Fqxy_MFCCA_0}
  F^0_{XY,\mathrm{MFCCA}}(s) &= \frac{1}{2N_s} \sum_{v=1}^{2N_s}\mathrm{sgn}\left [ f_{XY}^2(s, v)\right ] \ln \left | f_{XY}^2(s, v) \right |,
\end{align}
for $q=0$, where $\mathrm{sgn}$ is the sign function.

Another method that is frequently discussed in the research field is the MFDCCA~\cite{Zhou2008}, in which the absolute value of detrended components is introduced before the process of taking the generalized average,
\begin{align}
  \label{eq:fxy_abs}
  f_{XY, \mathrm{abs}}^2(s, v) = \frac{1}{s} \sum_{t=1}^{s}\left |X_v(t)-\tilde{X}_v(t)\right | \left |Y_v(t)-\tilde{Y}_v(t)\right |.
\end{align}
At any value of $q$, the $q$th-order fluctuation function can be calculated without the occurrence of an imaginary part.
\begin{align}
  \label{eq:Fqxy_MFDCCA}
  F^q_{XY,\mathrm{MFDCCA}}(s) = \frac{1}{2N_s} \sum_{v=1}^{2N_s}\left \{ f_{XY, \mathrm{abs}}^2(s, v) \right \}^{q/2}.
\end{align}
For $q=0$, the function is defined by the standard logarithmic averaging form~\cite{Kantelhardt2002},
\begin{align}
  \label{eq:F0xy_log_average}
  F^0_{XY,\mathrm{MFDCCA}}(s) = \exp \left[ \frac{1}{4N_s}\sum_{v=1}^{2N_s}\ln \left|f_{XY,\mathrm{abs}}^2(s, v) \right| \right].
\end{align}

By repeating the process for different $s$, the scale-dependent relationship between $F^q_{XY,A}(s)$ and $s$ is revealed, where $A \in \{\mathrm{MFCCA}, \mathrm{MFDCCA}\}$.
For simplicity, in this study, the functions based on MFCCA and MFDCCA are called the MFCCA function and MFDCCA function, respectively.
A power-law relationship of $\left [F^q_{XY,A}(s) \right ]^{1/q} \sim s^{h_{XY,A}(q)}$ appears when there exist long-range cross-correlations
\footnote{For $q=0$, the corresponding logarithmic averaging form is used. Also, if the MFCCA function is negative over the whole scaling range, its sign is reversed before estimating the scaling exponent.}.
The scaling exponent $h_{XY,A}(q)$ corresponds to the generalized Hurst exponent, which quantitatively describes multifractal properties of the power-law cross-covariance under algorithm $A$; in the notation of O{\'s}wi{\k{e}}cimka et al.~\cite{Owiecimka2014} this bivariate exponent is denoted $\lambda_q$, and for $X=Y$ it reduces to the ordinary generalized Hurst exponent $h(q)$. The exponents indicate whether the cross-correlations of series present long- or short-memory, in addition to whether they present multifractal structure~\cite{Kantelhardt2002, Thompson2016}. Parameter $q$ indicates the different magnitudes of fluctuations. If $q>0$, mainly the large level of fluctuations are amplified dominating the overall behavior, and if $q<0$, mainly the small level of fluctuations dominate the overall behavior. For the univariate case $X=Y$, $h_{XX,A}(2)$ corresponds to the standard Hurst exponent for a stationary series, while $h_{XY,A}(2)$ is its bivariate counterpart for two series.
When $h_{XY,A}(q)$ remains constant, the underlying process is monofractal, otherwise multifractal.

When $h_{XY,A}(q)$ is needed, it is obtained by a log--log regression over a finite scaling range. The lower and upper cutoffs should be selected to avoid polynomial-detrending artifacts at very small scales and finite-size bias at very large scales~\cite{Kantelhardt2002, Owiecimka2014, Thompson2016}. A reasonable scaling range of $20\leq s \leq N/5$ is selected in this study where $N$ is the length of the series~\cite{Owiecimka2014, Thompson2016}. This range is used for estimating scaling exponents. When $X=Y$, both MFCCA and MFDCCA reduce to the classical MFDFA algorithm~\cite{Kantelhardt2002}, so the analysis is simply on the auto-correlation of a single series.

\subsection{\label{sec:Mean-MMFC}The mean-MFCCA portfolio model}
In the traditional mean-variance model~\cite{Markowitz1952}, the asset allocation that minimizes variance of portfolio returns under a given expected return is selected. For $n$ assets, the variance is constructed using covariance $\sigma_{ij}$ among assets and expected returns $E(r_i)$ of individual assets. With the constraint that the process of minimization is achieved under some given return $r_e$, the investment weight $w_i$ for each asset can be obtained by solving the following optimization problem
\begin{equation}
  \label{eq:port_meanvariance}
  \begin{aligned}
    & \underset{w}{\text{minimize}}
    & & \frac{1}{2}\sum_{i,j=1}^{n}w_i w_j \sigma_{ij} \\
    & \text{subject to}
    & & \sum_{i=1}^{n} w_i E(r_i) \geq r_e, \; \sum_{i=1}^{n} w_i = 1, \\
    & & & w_i \geq 0, \; i = 1, \ldots, n,\\
  \end{aligned}
\end{equation}
where the objective function represents the variance of portfolio returns, while the no-short-selling constraint is imposed. The required-return condition is imposed as an inequality, so that the portfolio is required to attain at least the level $r_e$; this keeps the problem feasible whenever some asset attains the required return. Optimal allocations are estimated under linear correlations, so the model does not take into account the fractal characteristics of assets.

Building on an alternative measure of portfolio risk, the framework of the mean-DCCA model was developed to highlight fractal correlations and their underlying multiscale properties between two assets~\cite{CHUN2020}. In particular, the DCCA function was introduced in place of variance to represent portfolio risk, and this approach was applied to multiple-asset allocations~\cite{Zhang2022}. The mean-DCCA model was then generalized to the mean-MFDCCA-based criterion to take multifractal correlations into consideration by employing the MFDCCA method~\cite{Li2021MeanMFDCCA, Li2024MultiAssetMD}.

The present model extends this line of work by using MFCCA as the portfolio risk criterion, so that the sign of local detrended covariance is retained when portfolio risk is evaluated. The distinction from MFDCCA-based criteria matters when positive and negative local co-movements coexist across scales or fluctuation orders, because diversification depends on the signed balance of local interactions as well as on the strength of co-fluctuation.

We consider a portfolio of $n$ assets in which the conventional covariance matrix is replaced by the signed, scale-dependent, and fluctuation-order-dependent MFCCA fluctuation matrix. For the mean-MFCCA optimization problem, $F^q_{ij,\mathrm{MFCCA}}(s)$ is used as the $(i, j)$ element of the matrix. The portfolio is solved through the minimization process with the constraint shown as follows:
\begin{equation}
  \label{eq:port_fractal}
  \begin{aligned}
    & \underset{w}{\text{minimize}}
    & & \frac{1}{2}\sum_{i,j=1}^{n}w_i(q, s)w_j(q, s)F^q_{ij,\mathrm{MFCCA}}(s) \\
    & \text{subject to}
    & & \sum_{i=1}^{n} w_i(q, s) E(r_i) \geq r_e, \; \sum_{i=1}^{n} w_i(q, s) = 1, \\
    & & & w_i(q, s) \geq 0, \; i = 1, \ldots, n.\\
  \end{aligned}
\end{equation}
where short selling is not allowed, and the portfolio is required to realize risk minimization while attaining at least the given return level $r_e$. Investment weights $w_i(q, s)$ now depend on two parameters: the scale $s$ and the fluctuation order $q$. When $q=2$, the MFCCA fluctuation matrix corresponds to the DCCA covariance-type fluctuation matrix used in the mean-DCCA-based model, and the investment weights depend only on $s$. Otherwise, the objective function should be interpreted as a $q$th-order multifractal risk measure rather than the usual variance, and weights are determined under a pre-given pair of $(q, s)$.

A remark on the interpretation of this criterion is in order. For $q=2$, the local detrended covariance in Eq.~\ref{eq:fxy} is bilinear in the profiles, so the quadratic form $\sum_{i,j}w_iw_jF^2_{ij,\mathrm{MFCCA}}(s)$ coincides with the DCCA fluctuation function of the portfolio series itself, and the criterion is a direct scale-dependent analogue of the portfolio variance. For $q\neq2$, however, the power transformation applied within each segment breaks this additivity, and the quadratic form no longer equals the $q$th-order fluctuation function of the portfolio series. In this case, Eq.~\ref{eq:port_fractal} should be understood as a $q$th-order risk criterion that aggregates signed pairwise dependence information at the chosen scale and fluctuation order, rather than as the portfolio's own multifractal fluctuation measure. The same caveat applies to mean-MFDCCA-type criteria~\cite{Li2021MeanMFDCCA, Li2024MultiAssetMD}, which are built from pairwise MFDCCA functions in the same manner.

If the no-short-sale constraints are inactive, the return constraint binds, and the fluctuation matrix is nonsingular, the Lagrange multiplier method gives the equality-constrained analytical reference solution of $w_i(q, s)$, writing $r_i$ for $E(r_i)$, as follows:

\begin{equation}
  \label{eq:Lagrange_multiplier_method}
  \begin{gathered}
    w_i(q,s)
    =
    \frac{(C-Br_e)a_i+(Ar_e-B)b_i}{D},
    \\
    a_i=\sum_{j=1}^{n}\theta_{ij},\quad
    b_i=\sum_{j=1}^{n}\theta_{ij}r_j,
    \\
    A=\sum_{k,l=1}^{n}\theta_{kl},\quad
    B=\sum_{k,l=1}^{n}\theta_{kl}r_l,
    \\
    C=\sum_{k,l=1}^{n}r_k\theta_{kl}r_l,\quad
    D=AC-B^2,
    \\
    \Theta=[\theta_{ij}]_{i,j=1}^{n}=\left(\mathbf{F}_{\mathrm{MFCCA}}^q(s)\right)^{-1},
    \\
    \mathbf{F}_{\mathrm{MFCCA}}^q(s)=\left[F_{ij,\mathrm{MFCCA}}^q(s)\right]_{i,j=1}^{n}.
  \end{gathered}
\end{equation}
The corresponding mean-MFDCCA benchmark is defined by solving the same constrained optimization problem after replacing $F^q_{ij,\mathrm{MFCCA}}(s)$ in Eq.~\ref{eq:port_fractal} and, for the equality-constrained reference solution, Eq.~\ref{eq:Lagrange_multiplier_method} with $F^q_{ij,\mathrm{MFDCCA}}(s)$.

When $\mathbf{F}_{\mathrm{MFCCA}}^q(s)$ is positive semidefinite (PSD), the associated quadratic form defines a non-negative multifractal risk criterion. Although sign-preserving cross-fluctuation terms do not guarantee PSD by construction, non-PSD cases are uncommon in typical applications because the positive auto-fluctuation terms on the diagonal tend to dominate the cross-fluctuation terms, and the averaging over local segments stabilizes the estimated matrix. If finite-sample estimation nevertheless yields a non-PSD matrix, we regularize it to the nearest PSD matrix before portfolio optimization, so the portfolio problem becomes a convex quadratic program under the long-only constraints~\footnote{Specifically, the estimated MFCCA fluctuation matrix is first symmetrized, and negative eigenvalues are replaced with zero by eigenvalue clipping to obtain a positive semidefinite approximation. In the numerical experiments in this study, all MFCCA fluctuation matrices were PSD, so no regularization was required. In the empirical analysis, non-PSD cases were rare; see Section IV for details.}.

Given that the FMH assumes that the market is composed of diverse types of investors operating across various investment horizons, the influence of multiple temporal scales must be incorporated into the determination of optimal weights~\cite{Peters1994, Zhang2022, Li2024MultiAssetMD, Kakinaka2023}. The weight $w_i(q, s)$ derived from a specific time horizon $s$ and a specific fluctuation order $q$ illustrates merely a fractional component of the complex multiscale dynamics of the market and is insufficient for detecting the optimal investment weight in an overall context. Thus, additional elaborations are necessary to ensure the portfolio's effectiveness.
Building upon prior research~\cite{Kakinaka2023}, we consider a set of time scales $S=\left \{ s_{\mathrm{min}}, \ldots, s_{\mathrm{max}} \right \}$, where $s_{\mathrm{min}}$ and $s_{\mathrm{max}}$ are the minimum and maximum elements of set $S$. We also account for multifractal effects by regarding a range of fluctuation orders $Q=\left \{ q_{\mathrm{min}}, \ldots, q_{\mathrm{max}} \right \}$. Both the multiple time scales and the fluctuation orders should be integrated into the investment weight. One way is to define the optimal weight $w^{\mathrm{opt}}_i$ as the weighted average of $w_i(q, s)$ with scales that satisfy $s\in S$ and $q \in Q$.
\begin{equation}
  \label{eq:optimal_weight}
  \begin{aligned}
    w^{\mathrm{opt}}_i = \sum_{s\in S} \sum_{q\in Q} \alpha(s) \beta(q) w_i(q, s).
  \end{aligned}
\end{equation}
For the weight budget, $\sum_{s\in S} \alpha(s)=1$ holds where $\alpha(s) \in [0, 1]$ represents the investor's relative preference degree for time scale $s$, and $\sum_{q\in Q} \beta(q)=1$ holds where $\beta(q) \in [0, 1]$ represents the investor's relative preference degree for fluctuation amplitudes.
Equation~\ref{eq:optimal_weight} is not intended to minimize a single aggregated risk function. Rather, $w_i(q,s)$ is interpreted as the optimal allocation of an investor type characterized by scale $s$ and fluctuation-order sensitivity $q$, while $\alpha(s)\beta(q)$ represents the relative importance, or preference mass, of that type in the heterogeneous investor population. Thus, Eq.~\ref{eq:optimal_weight} is a population-level aggregation of investor-type-specific optimal allocations under heterogeneous investment horizons and fluctuation sensitivities.

The preference profiles $\alpha(s)$ and $\beta(q)$ can encode different investor populations, such as the short- and long-horizon preferences considered in Ref.~\cite{Kakinaka2023}. For simplicity, we assume no specific preference across horizons or fluctuation magnitudes and set $\alpha(s)=\frac{1}{\# S}$ and $\beta(q)=\frac{1}{\# Q}$, where $\#$ denotes the number of elements of the set; this leads to a \textit{naive} allocation in the fractal portfolio setting. The sets $S$ and $Q$ and the profiles $\alpha(s)$ and $\beta(q)$ are chosen by the investor and are not estimated from the data. How much the empirical results depend on these choices is examined in Sec.~\ref{sec:robustness}.

Throughout the remainder of the paper we use the following abbreviations. MV denotes the conventional mean--variance portfolio, while MMFC and MMFD denote the mean-MFCCA and mean-MFDCCA portfolios aggregated over $S$ and $Q$ as in Eq.~\ref{eq:optimal_weight}. Their single-order $Q=\{2\}$ versions are denoted MC and MD, respectively. Since MFCCA reduces to DCCA at $q=2$, MC coincides with the mean-DCCA model of Refs.~\cite{CHUN2020, Zhang2022}, whereas MD is its absolute-value (MFDCCA-based) counterpart.

\section{Numerical Experiments with Computer Generated Series}
To confirm that the proposed mean-MFCCA provides effective risk diversification, we check the portfolio's performance under two types of simulated time series.
\subsection{Multiscale diversification}\label{subsec:multiscale}
For the purpose of testing whether the proposed mean-MFCCA framework reflects different diversification effects across scales, we first generate synthetic series based on the two-component autoregressive fractionally integrated moving average (ARFIMA) stochastic process. This process is designed to generate two coupled fractal signals with long-range power-law auto-correlations and cross-correlations~\citep{Podobnik2008TwoComponentARFIMA, podobnik2009quantifying, zebende2011dcca}. The generated series $y_t$ and $y'_t$ are defined as
\begin{equation}
  \label{eq:twocomponentARFIMA}
  \begin{aligned}
    y_t = W \sum_{j=1}^\infty a_j(d_1)y_{t-j} +(1-W) \sum_{j=1}^\infty a_j(d_2)y'_{t-j} + \varepsilon_t \\
    y'_t = (1-W) \sum_{j=1}^\infty a_j(d_1)y_{t-j} +W \sum_{j=1}^\infty a_j(d_2)y'_{t-j} + \varepsilon^{\prime}_t.
  \end{aligned}
\end{equation}
Here, $a_j(d)$ represents statistical weights, which is defined using the Gamma function $\Gamma$ and the exponent parameter $d\in (-0.5, 0.5)$,
\begin{align*}
  a_j(d) = \frac{\Gamma(j-d)}{\Gamma(-d)\Gamma(1+j)}.
\end{align*}
Note that $d$ is related to the Hurst exponent in terms of DFA analysis, where $H=d+0.5$ holds for the stationary ARFIMA range $d\in(-0.5,0.5)$~\citep{Podobnik2008TwoComponentARFIMA, podobnik2009quantifying, zebende2011dcca}.
The residuals $\varepsilon_t$ and $\varepsilon'_t$ are independent and identically distributed Gaussian variables with zero mean and unit variance.
The parameter $W\in[0.5,1]$ controls the strength of power-law cross-correlations between $y_t$ and $y'_t$ in the long-term: $W=1$ corresponds to the weakest coupling, whereas $W=0.5$ corresponds to the strongest symmetric coupling. The innovation structure determines the short-term dependence; for example, identical innovations produce strong short-term co-movement, opposite innovations produce short-term anti-correlations, and independent innovations imply weak short-term correlations.

\begin{figure*}[!t]
  \centering
  \begin{tabular}{@{}cc@{}}
    \begin{minipage}{0.44\linewidth}
      \centering
      \includegraphics[width=\linewidth]{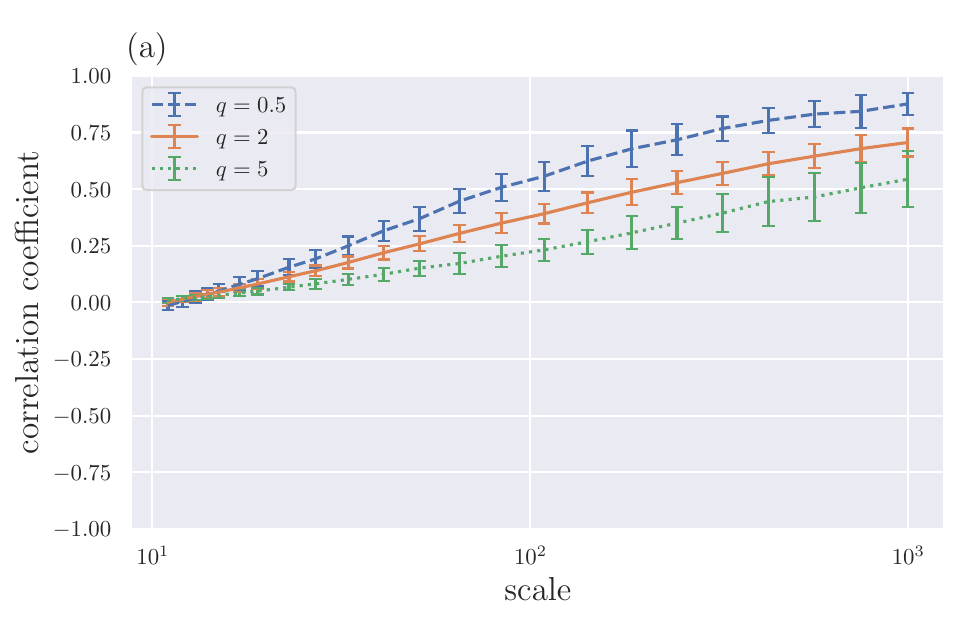}
    \end{minipage}
    \begin{minipage}{0.44\linewidth}
      \centering
      \includegraphics[width=\linewidth]{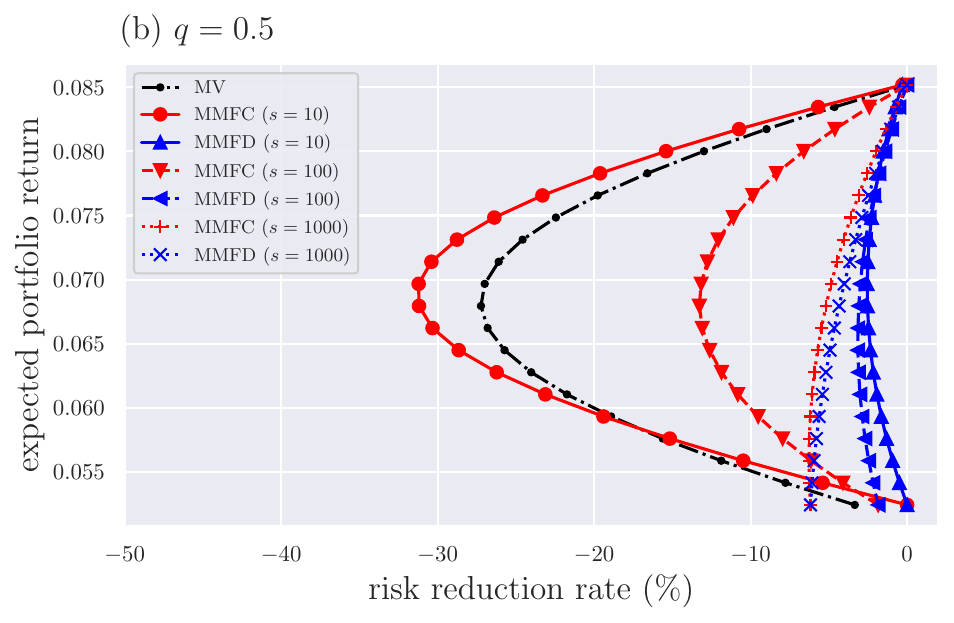}
    \end{minipage} \\
    \begin{minipage}{0.44\linewidth}
      \centering
      \includegraphics[width=\linewidth]{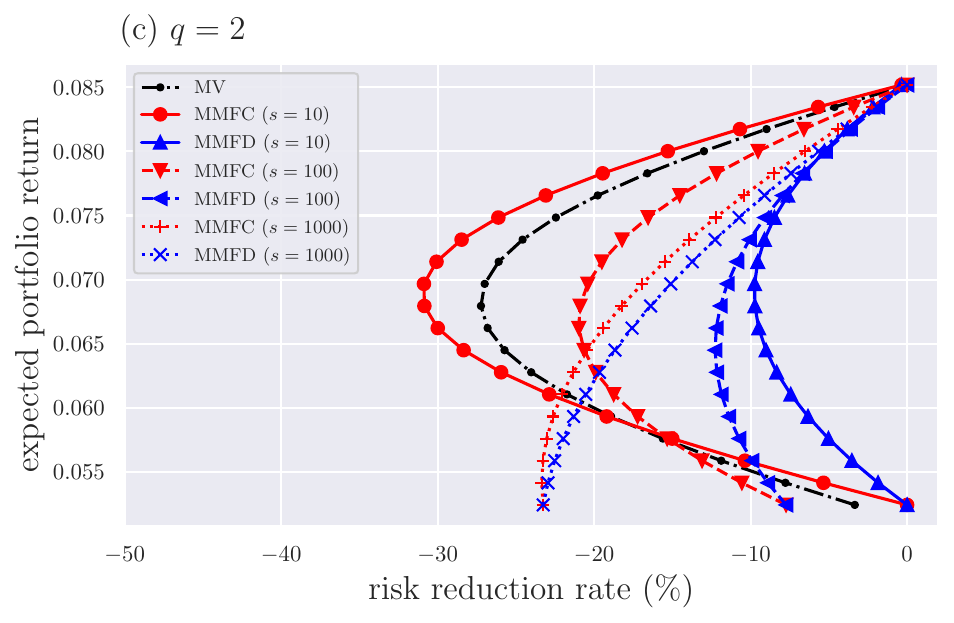}
    \end{minipage}
    \begin{minipage}{0.44\linewidth}
      \centering
      \includegraphics[width=\linewidth]{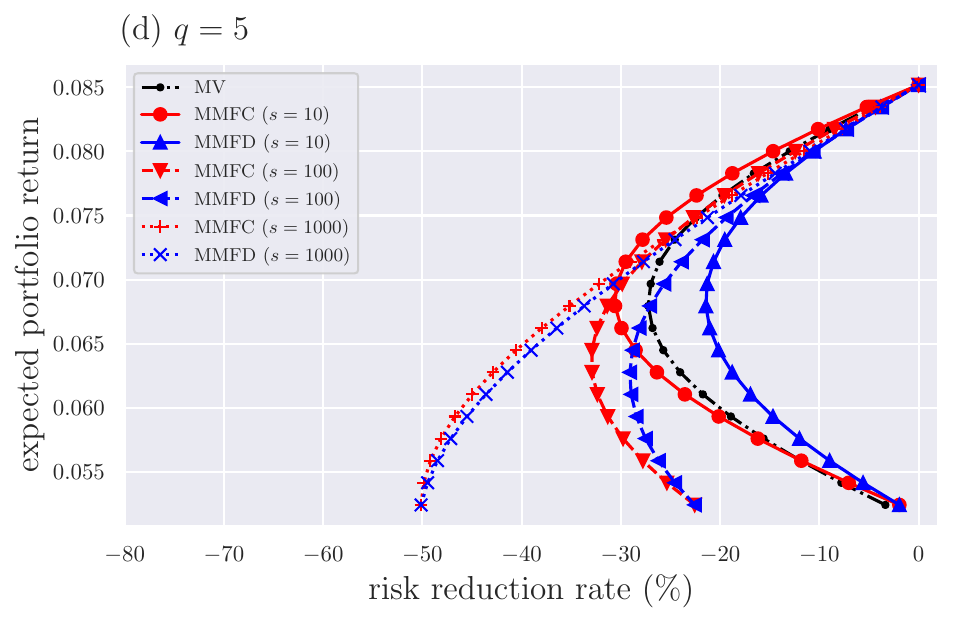}
    \end{minipage}
  \end{tabular}
  \caption{(a) Cross-correlations between scale-dependent simulated series generated by the two-component ARFIMA model. The coefficients are weak at short scales because the innovations are independent $(\varepsilon_t \neq \varepsilon'_t)$, but they gradually increase as the scale becomes larger because of long-range power-law cross-correlations.
    (b) (c) (d) Efficient frontiers of the two types of fractal portfolio among different scales $(s=10, 100, 1000)$.
  Each portfolio is observed under different fluctuation orders, $(b): q=0.5$, $(c): q=2$, and $(d): q=5$, to check robustness. The fluctuation order is shown in each panel label. The horizontal axis is the risk reduction rate defined in Eq.~\ref{eq:rrr}, which measures the shape of the frontier. For all $q$ values, the frontier shape varies with $s$, indicating different levels of diversification effects across scales.}
  \label{fig:twoComARFIMA_efficientfrontier}
\end{figure*}

We generate series with $d_1=0.4$, $d_2=0.2$, $W=0.7$, and independent innovations $\varepsilon_t \neq \varepsilon'_t$ of length $2^{13}$.
This setting produces a pair of series in which short-term correlations are weak because of independent innovations, whereas positive power-law cross-correlations become stronger at longer time scales.
For the two generated series, we calculate the scale-dependent cross-correlation levels by using the $q$DCCA coefficient of \cite{Kwapie2015},
\begin{equation}
  \label{eq:coef_qdcca}
  \begin{aligned}
    \rho_{q\mathrm{DCCA}}(q, s)=\frac{F^q_{XY,A}(s)}{\sqrt{F^q_{XX,A}(s) F^q_{YY,A}(s)}},
  \end{aligned}
\end{equation}
where $A$ denotes the fluctuation-function algorithm used in the calculation, and $q$ is the fluctuation order. Note that when $q=2$, the $q$DCCA coefficient reduces to the ordinary DCCA-type coefficient of \cite{zebende2011dcca}.
The cross-correlations appear to be scale-dependent: they are weak or almost absent at short time scales and become stronger at longer time scales (Fig.~\ref{fig:twoComARFIMA_efficientfrontier}a). This confirms that the simulated pair is suitable for examining multiscale diversification effects.
The main scale-dependent pattern remains robust across the selected $q$ values.

Next, the effect of multiscale properties on portfolio diversification when applying the mean-MFCCA/MFDCCA portfolio is investigated.
Which portfolio is more efficient depends on how risk is measured, and comparing their performance does not make sense when risks have different definitions~\cite{Cheng2001}. We address this point by observing the \textit{efficient frontier}, a standard concept in finance that represents a set of optimal investment portfolios offering the lowest possible risk for a given level of return.
The risk criteria of the models have different units, so the frontiers cannot be drawn on a common risk axis. We therefore plot each frontier against its own \textit{risk reduction rate}. Let $\mathbf{M}$ be the risk matrix of a model, that is, the covariance matrix for MV and $\mathbf{F}^q_{A}(s)$ for the fractal models, and let $w^{*}(r_e)$ be the frontier portfolio at required return $r_e$. Writing $\mathcal{R}(w)=\sqrt{w^{\top}\mathbf{M}w}$ for the square root of the minimized quadratic form, we define
\begin{equation}
  \label{eq:rrr}
  \mathrm{RR}(r_e)=100\times\frac{\mathcal{R}\!\left(w^{*}(r_e)\right)-\mathcal{R}_{\max}}{\mathcal{R}_{\max}},
\end{equation}
where $\mathcal{R}_{\max}=\max_{r}\mathcal{R}(w^{*}(r))$ is the largest value of $\mathcal{R}$ on the frontier. Since the frontier runs between the two single-asset portfolios, $\mathcal{R}_{\max}$ is attained at one of them, namely at the asset that is riskier under the criterion in question. The rate is thus zero for that undiversified portfolio and negative elsewhere, and a value of $-30\%$ means that diversification lowers the model's own risk criterion by 30\% relative to it. Because $\mathcal{R}$ is defined through each model's own matrix, the rates of two different models are not comparable as risk levels, and only the shapes of the frontiers are compared.
Since the data-generating process is constructed so that the cross-correlation structure changes explicitly with scale $s$, changes in the shape of the efficient frontier provide a mechanistic validation that the proposed portfolio criterion transmits multiscale dependence information into allocation outcomes.
Therefore, the purpose of this comparison is not to prove uniform risk dominance, but to verify that the controlled dependence structure is reflected in the portfolio frontier.

Comparing the efficient frontiers with scale subsets $S = \{10\}, \{100\}, \{1000\}$ under various fluctuation order subsets $Q=\{0.5\}, \{2\}, \{5\}$, we confirm that the curve of the efficient frontier as depicted differs among scales (Fig.~\ref{fig:twoComARFIMA_efficientfrontier}).
The result indicates that the scale dependence of the serial correlations and cross-correlations can be captured in the portfolio in a more precise manner. As the scale increases from $s=10$ to $s=1000$, the efficient frontier approaches a straighter shape, which is consistent with the fact that the simulated series $y_t$ and $y'_t$ lose diversification power because of stronger correlations at larger scales. This trend is also observed when $q$ is set to other values.
We also confirm that in this controlled setting, the MFCCA-based frontier shows a larger deformation toward lower risk levels compared to the MFDCCA-based frontier, suggesting that the model is more sensitive to the designed scale-dependent dependence structure.

\subsection{Multifractal diversification}\label{subsec:multifractal}

\begin{figure*}[!t]
  \centering
  \begin{tabular}{@{}cc@{}}
    \begin{minipage}{0.44\linewidth}
      \centering
      \includegraphics[width=\linewidth]{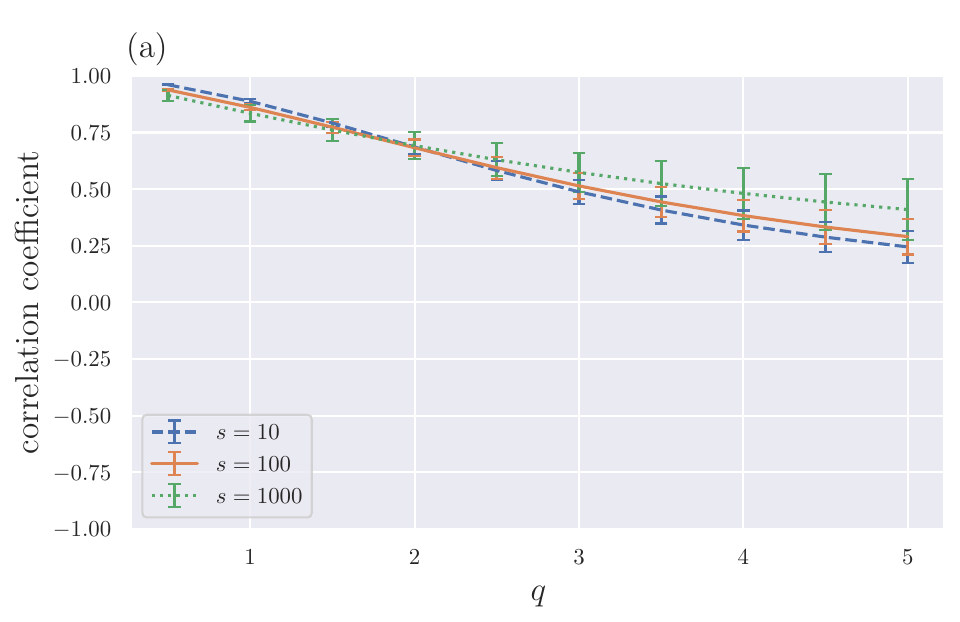}
    \end{minipage}
    \begin{minipage}{0.44\linewidth}
      \centering
      \includegraphics[width=\linewidth]{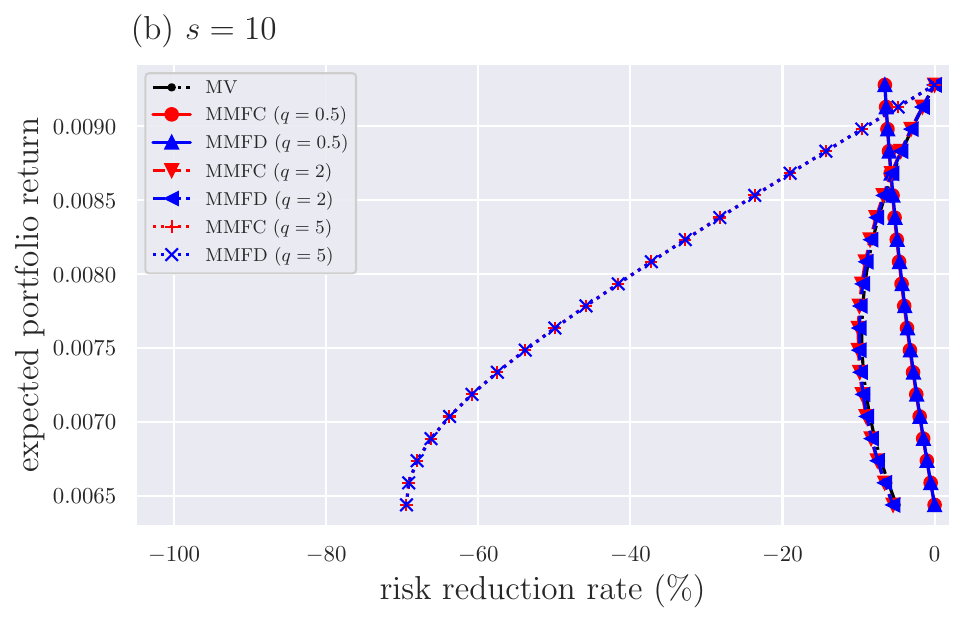}
    \end{minipage} \\
    \begin{minipage}{0.44\linewidth}
      \centering
      \includegraphics[width=\linewidth]{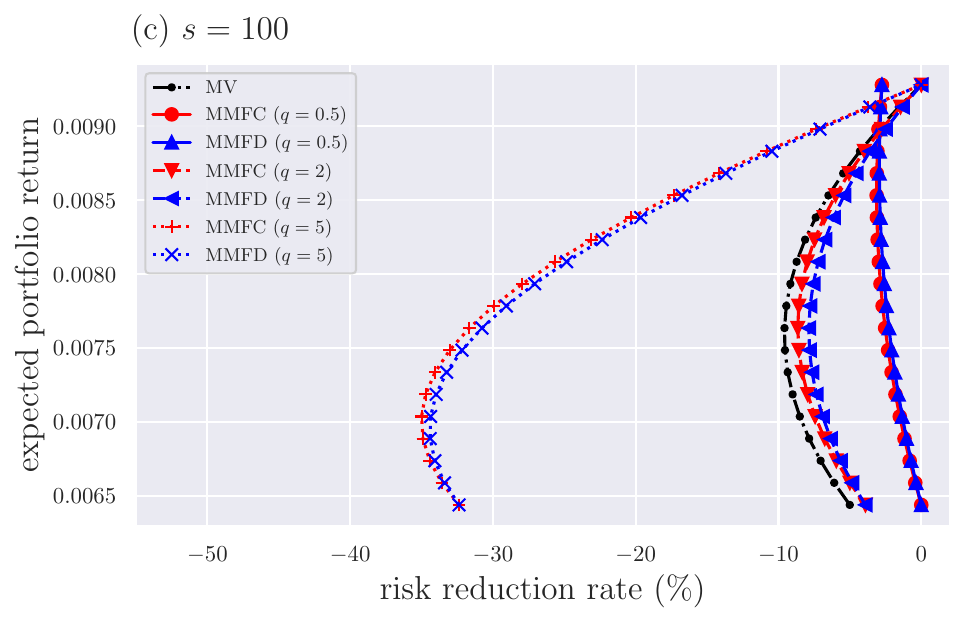}
    \end{minipage}
    \begin{minipage}{0.44\linewidth}
      \centering
      \includegraphics[width=\linewidth]{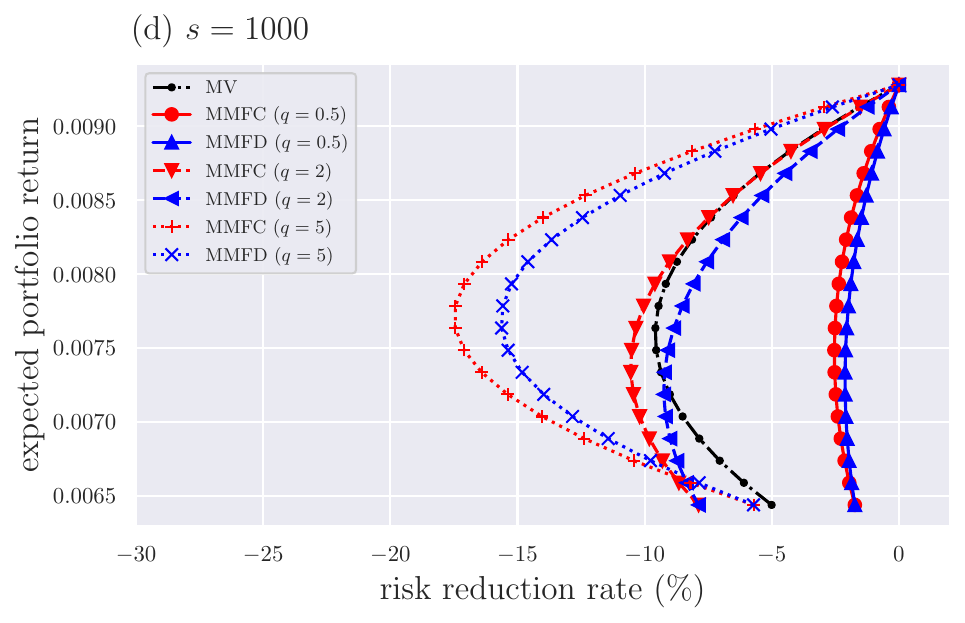}
    \end{minipage}
  \end{tabular}
  \caption{(a) Cross-correlations between multifractal simulated series generated by the Markov-switching multifractal model. The $q$DCCA coefficients are calculated at specific scales $s=10, 100, 1000$. The coefficients are strongly positive for small fluctuations, whereas they remain positive but become weaker for large fluctuations.
    (b) (c) (d) Efficient frontiers of the two types of fractal portfolio among different $q$-th order fluctuations $(q=0.5, 2, 5)$.
    Each portfolio is observed based on different scales, $(b): s=10$, $(c): s=100$, and $(d): s=1000$, to check robustness. The scale is shown in each panel label. The horizontal axis is the risk reduction rate defined in Eq.~\ref{eq:rrr}, which measures the shape of the frontier. For all scales, the frontier shape varies with $q$, indicating different levels of diversification effects among fluctuation orders.
  }
  \label{fig:MSM_efficientfrontier}
\end{figure*}

In order to examine whether the MMFC portfolio can yield diversification effects that reflect multifractal characteristics, we next utilize the Markov-switching multifractal model (MSM) and generate two series with multifractal scaling and multifractal cross-correlations~\citep{Owiecimka2014}. MSM is suitable for this purpose because it represents volatility as a hierarchy of randomly renewed multipliers, thereby generating intermittent fluctuations, volatility clustering, and multiscaling behavior similar to those observed in financial returns~\citep{Liu2007MSMLongRange, Kwapien2005ComponentsMultifractality}. In particular, MSM can reproduce apparent long-memory scaling in financial volatility, while empirical evidence from high-frequency stock returns indicates that multifractality is associated with both fat-tailed distributions and nonlinear temporal correlations~\citep{Liu2007MSMLongRange, Kwapien2005ComponentsMultifractality}. Following the MSM construction, the observable $r_t^{(z)}$ for $z\in\{x,y\}$ is written as
\begin{equation}
  \label{eq:MSM_model}
  \begin{aligned}
    r_t^{(z)} &= \sigma_t^{(z)}u_t^{(z)}, \\
    \left(\sigma_t^{(z)}\right)^2 &= \sigma^2\prod_{i=1}^{k}M_i^{(z)}(t), \\
    \gamma_i &= 1-\left(1-\gamma_{k}\right)^{b^{i-k}},\quad i=1,2,\ldots,k,
  \end{aligned}
\end{equation}
where $u_t$ is a Gaussian innovation, $\sigma_t^{(z)}$ is the instantaneous volatility, and $M_i^{(z)}(t)$ denotes the $i$th volatility multiplier. At each time step, $M_i^{(z)}(t)$ is renewed with probability $\gamma_i$ and otherwise remains unchanged. When it is renewed, the multiplier is drawn from the binomial set $\{m_z,2-m_z\}$.

In our simulation, we set the series length to $N=2^{13}$ and use $m_x=1.2$ for $x_t$, $m_y=1.5$ for $y_t$, $k=10$, $b=2$, $\gamma_{k}=0.5$, and $\sigma=1$. The two series share the same renewal structure of the multipliers, and the same Gaussian innovation $u_t$ is used when signed MSM returns are generated. This setting follows the MSM-based multifractal cross-correlation experiment of O{\'s}wie{\c{c}}imka et al.~\citep{Owiecimka2014} and induces a hierarchical volatility dependence between $x_t$ and $y_t$ while allowing their multifractal strengths to differ.
For the two generated series, we calculate the cross-correlation levels at different fluctuation orders $q$ using the $q$DCCA coefficient of \cite{Kwapie2015} as shown in Fig.~\ref{fig:MSM_efficientfrontier}a.
The cross-correlation is studied under scales of $s=10, 100, 1000$. For all selected scales, the coefficients show signs of fluctuation-order dependence: they are strongly positive for small $q$, whereas they remain positive but become weaker for large $q$, indicating that small and large fluctuations contribute differently to the cross-correlation structure.

The MMFC, MMFD, and traditional MV portfolios are then constructed using the generated series.
As in the multiscale experiment, variation of the frontier across $q$ indicates that the portfolio criterion responds to the designed fluctuation-order-dependent structure.
Comparison of the efficient frontiers with fluctuation order subsets $Q=\{0.5\}, \{2\}, \{5\}$ under various scale subsets $S=\{10\}, \{100\}, \{1000\}$, shows that the frontier curve differs across $q$ (Fig.~\ref{fig:MSM_efficientfrontier}). It approaches a straighter line as $q$ increases, which is consistent with the weaker positive cross-correlations observed for larger fluctuations. The results imply that multifractal characteristics in the serial and cross-correlations of the series can be reflected in portfolio allocation.

\subsection{Portfolio performance under multiscale and multifractal diversification}
The next step in the analysis is to evaluate portfolio performance when both multiscale and multifractal diversification effects are present.
A pair of synthetic series $x^{\mathrm{twoC}}_t$ and $y^{\mathrm{twoC}}_t$ is generated from the two-component ARFIMA model with the same parameter settings as in subsection~\ref{subsec:multiscale}, and another pair of synthetic series $x^{\mathrm{MSM}}_t$ and $y^{\mathrm{MSM}}_t$ is generated from the Markov-switching multifractal model with the same parameter settings as in subsection~\ref{subsec:multifractal}.
The simulation procedure is repeated 50 times to reduce Monte Carlo variability, with a fixed random seed assigned to each realization so that all reported numbers are exactly reproducible.
For simplicity, we consider the minimum-risk portfolio by setting the required-return level low enough that the return constraint is inactive, so that effectively only the budget and no-short-sale constraints are imposed.

When assessing portfolio performance, we do not rely on variance-based indicators such as the Sharpe ratio, which are most meaningful under distributional or preference assumptions that fractal approaches deliberately relax~\cite{Yue2017}. We instead evaluate the portfolios by distribution-based downside measures that are defined independently of any of the competing risk criteria.
For this reason, we utilize Value-at-Risk (VaR) and Expected Shortfall (ES), which evaluate downside risk from the tail of the portfolio return distribution. To make the sign convention clear, let $R_p$ denote the portfolio return and $L_p=-R_p$ denote the portfolio loss. We report VaR and ES as positive loss measures:
\begin{align*}
  \mathrm{VaR}_{\alpha}(L_p) &= Q_{\alpha}(L_p) = -Q_{1-\alpha}(R_p), \\
  \mathrm{ES}_{\alpha}(L_p) &= \mathbb{E}\left[L_p \mid L_p \geq \mathrm{VaR}_{\alpha}(L_p)\right],
\end{align*}
where $Q_{\alpha}$ denotes the empirical $\alpha$-quantile. Following Basel~III market-risk practice, VaR and ES are calculated at the 99\% and 97.5\% confidence levels, respectively~\cite{BaselMAR33, Artzner1999, AcerbiTasche2002}. Both measures are computed from 10-period cumulative portfolio returns, where one period corresponds to one time step of the simulated series (interpreted as one trading day). The average values of these two loss measures are calculated over the 50 simulated portfolio realizations.
For the MMFD and MMFC, the scale subset and the fluctuation order subset of Eq.~\ref{eq:optimal_weight} are set to $S=\{10, 100, 1000\}$ and $Q=\{0.5, 1, 1.5,\ldots, 4.5, 5\}$, respectively, to calculate the optimal weight $w^{\mathrm{opt}}$.

In addition to MMFD and MMFC, their single-order $Q=\{2\}$ counterparts MD and MC are reported separately.

\begin{table}[htbp]
  \caption{\label{tab:simdata_performance}%
    Portfolio performance comparison using synthetic simulated series---one pair of series is generated from the two-component ARFIMA model and another pair is generated from the Markov-switching multifractal model, as explained in the previous subsections. Here, the scale subset is set as $S=\{10, 100, 1000\}$ and for the MMFD and MMFC the fluctuation order subset is set as $Q=\{0.5, 1, 1.5,\ldots, 4.5, 5\}$. The averages of 50 seeded realizations of 10-period 99\% VaR and 97.5\% ES are shown as positive loss measures. In parentheses, we show the standard deviations of the realizations. MD denotes the MFDCCA-type $q=2$ benchmark, and MC denotes the signed MFCCA-type $q=2$ benchmark.
  }
  \begin{ruledtabular}
    \begin{tabular}{lccccc}
      Portfolio & MV & MD & MMFD & MC & MMFC \\ \hline
      VaR99 & 6.677 & 6.342 & 6.193 & 5.925 & 5.923 \\
      & (0.722) & (0.542) & (0.570) & (0.614) & (0.607) \\
      ES97.5 & 6.742 & 6.603 & 6.433 & 6.095 & 6.048 \\
      & (0.733) & (0.597) & (0.590) & (0.612) & (0.594) \\ \hline
      Risk fn. & Cov. & MFDCCA & MFDCCA & MFCCA & MFCCA \\
      Orders & -- & $q=2$ & $q\in Q$ & $q=2$ & $q\in Q$ \\
    \end{tabular}
  \end{ruledtabular}
\end{table}

Table~\ref{tab:simdata_performance} provides a benchmark comparison and shows that fractal-based portfolios reduce downside losses relative to the MV benchmark.
Because VaR and ES are reported as positive loss measures, smaller values indicate stronger risk reduction. The largest losses occur under MV, followed in decreasing order by MD, MMFD, MC, and MMFC.
Since the five portfolios are evaluated on the same 50 realizations, the differences can be tested by paired comparisons. Wilcoxon signed-rank tests with Holm correction confirm that all pairwise differences in ES are statistically significant, including the improvement of MMFC over MC ($p<10^{-5}$); for VaR, all pairwise differences are significant except that between MC and MMFC, which are statistically indistinguishable at this sample size. The realized total returns of the five portfolios do not differ significantly across the realizations (Wilcoxon $p>0.3$ for all pairs), so the risk reduction is not obtained at the expense of returns.
This ranking suggests that incorporating scale-dependent and multifractal fluctuation structures into portfolio allocation can potentially improve downside-risk control.
The improvement of MMFC over MMFD stems from the insight that sign information in each detrended fluctuation subseries can play an important role in diversification. This finding is consistent with the observation in~\cite{Owiecimka2014} that fluctuation functions derived from MFCCA can quantify correlations between assets more accurately than those derived from MFDCCA. Notably, the $q=2$ sign-preserving benchmark MC already outperforms the fully multifractal but absolute-value-based MMFD. Sign preservation thus contributes more to downside-risk reduction than aggregation over fluctuation orders, and the additional contribution of the aggregation appears mainly in ES. Among the portfolios considered here, MMFC therefore shows the strongest tail-risk reduction.

Such improvement arises because the MSM series introduce fluctuation-order-dependent correlation structures, thereby producing multifractal diversification effects. Because ample evidence indicates that financial markets contain multiple scaling factors and exhibit multifractal correlations~\cite{Kantelhardt2002, Owiecimka2014, Kwapie2015, Thompson2016}, these effects are expected to carry over to the allocation of real assets, which we examine next.

\section{Application to empirical data}
To demonstrate the applicability of the MMFC portfolio to empirical research, we apply it to four representative financial return series: the Nikkei stock index, the S\&P 500 stock index, West Texas Intermediate (WTI) crude oil futures, and gold spot prices against the U.S. dollar (XAU/USD), covering the period from January 2011 to November 2023. Daily returns are calculated as log-price increments, using closing price data obtained from \url{http://www.histdata.com/}. To apply the MFCCA algorithm consistently, observations with missing values are removed so that all return series are aligned on common available trading dates, yielding $N=3330$ observations.
Daily observations are constructed from one-minute quotes by aggregating each series over a trading day defined as the interval between consecutive 17:00 Eastern-time trading pauses, so that each daily closing price corresponds to the last traded quote of the session. Since the provider distributes dealer quotes for the index and commodity series, the data should be regarded as tradable proxies of the underlying assets.

\subsection{Scaling properties of empirical data}
\begin{figure*}[!t]
  \centering
  \begin{tabular}{@{}cc@{}}
    \begin{minipage}{0.4\linewidth}
      \centering
      \includegraphics[width=\linewidth]{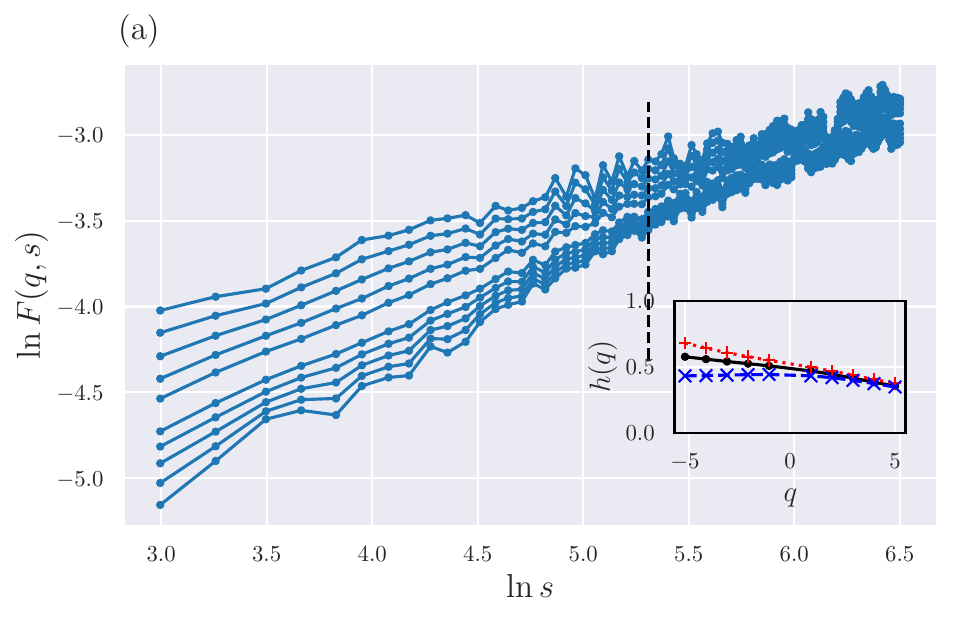}
    \end{minipage}
    \begin{minipage}{0.4\linewidth}
      \centering
      \includegraphics[width=\linewidth]{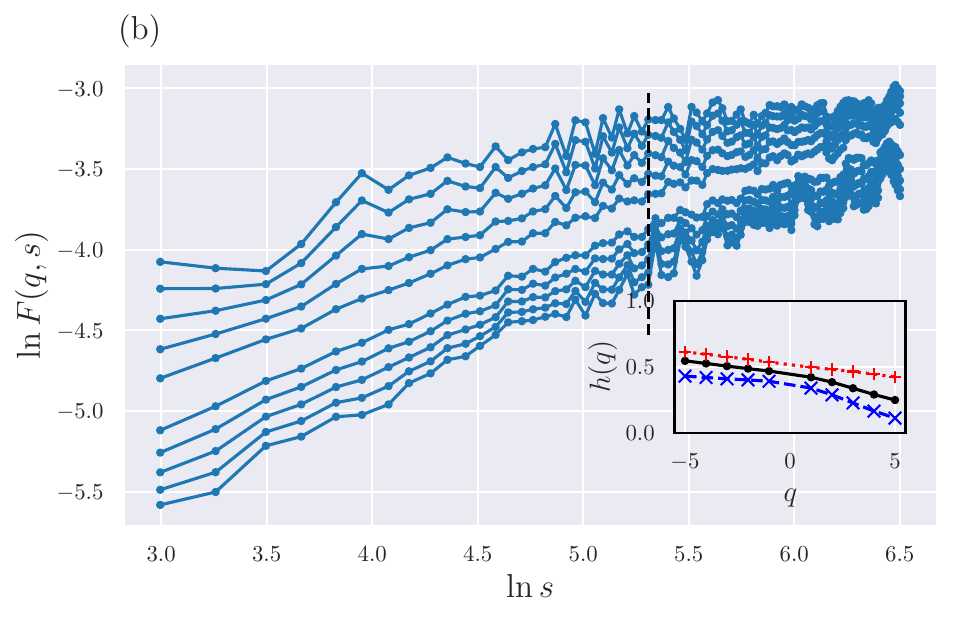}
    \end{minipage} \\
    \begin{minipage}{0.4\linewidth}
      \centering
      \includegraphics[width=\linewidth]{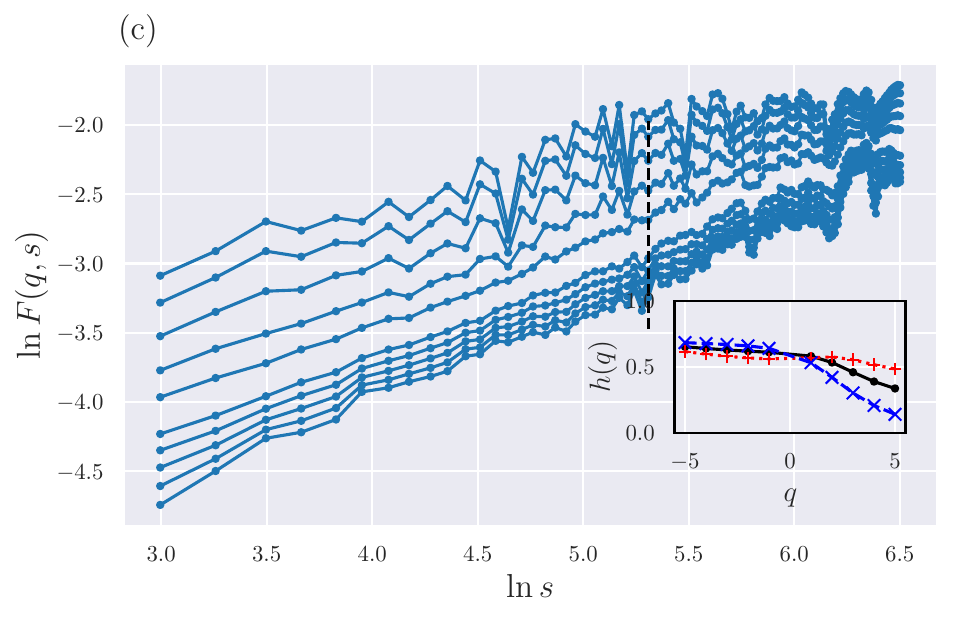}
    \end{minipage}
    \begin{minipage}{0.4\linewidth}
      \centering
      \includegraphics[width=\linewidth]{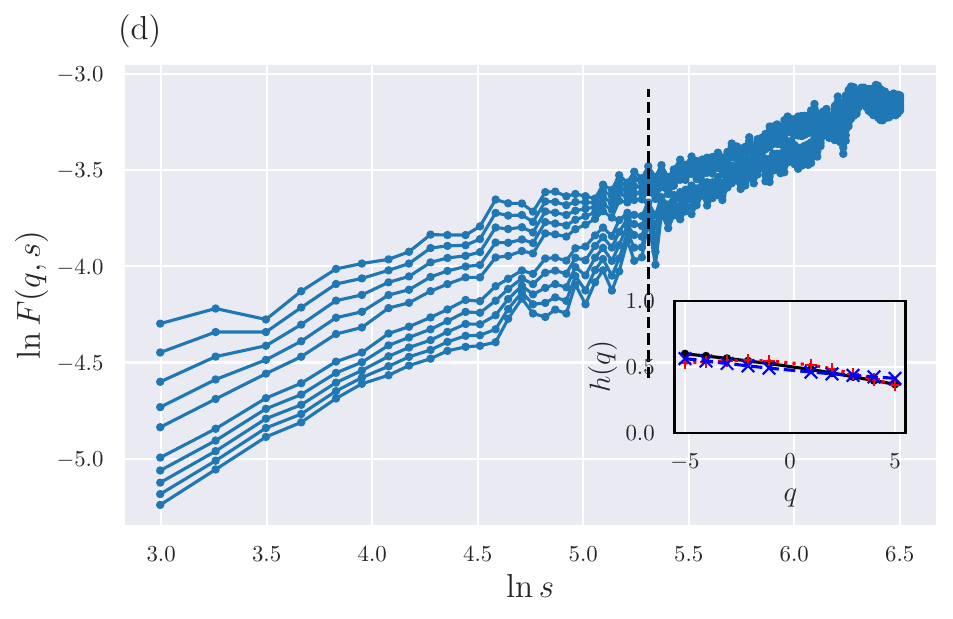}
    \end{minipage}
  \end{tabular}
  \caption{
    Scaling properties of empirical return series.
    The main panels show $\ln F(q,s)$ against $\ln s$ for (a) Nikkei, (b) S\&P500, (c) WTI, and (d) XAU.
  The dashed vertical line marks the common crossover scale $s_{\times}=200$. The inset panels show the generalized Hurst exponent $h(q)$ estimated over the full range $20\leq s\leq N/5$ (black circles, solid), over $20\leq s<s_{\times}$ (red plus signs, dotted), and over $s_{\times}\leq s\leq N/5$ (blue crosses, dashed).}
  \label{fig:empirical_autocorr}
\end{figure*}

We first examine the scaling properties of the empirical signals to assess whether the data are consistent with the FMH framework.
Figure~\ref{fig:empirical_autocorr} shows for each asset the logarithm of the detrended $q$-th order fluctuation function $F(q, s) = \left[F^q_{XX, MFCCA}(s)\right]^{1/q}$ plotted against the logarithm of the scale. A power-law pattern is observed in the autocorrelation for $q$ ranging from -5 to 5. Here, $F(q, s)$ takes a positive value for any value of $q$, because the correlation of individual time series is considered and $f^2(s,v)$ remains positive at all times.
A closer look at the power-law behavior shows that the slope is not constant over the full scale range. For the S\&P~500 and WTI, the curves for large $q$ flatten beyond $s_{\times}\simeq200$ trading days ($\ln s_{\times}\simeq5.3$), whereas the Nikkei and XAU show only mild changes of slope. A single power-law fit over the whole range may mix different regimes. We therefore fix the crossover scale to $s_{\times}=200$ for all assets and estimate $h(q)$ separately from the scales below and above $s_{\times}$. The insets of Fig.~\ref{fig:empirical_autocorr} show both estimates together with the global fit.
Within each range, $h(q)$ decreases with $q$ for every asset, which is the signature of multifractality. The above results are mostly consistent with recent empirical studies that document multifractal behavior or multifractal cross-correlations in stock and commodity-related markets using related complexity measures, such as scaling exponents and singularity spectra~\citep{Chen2024ChinaUSMFDCCA, Acikgoz2024GreenBondCommodityMFDCCA, Acikgoz2025EmergingFinancialMFDCCA}.

\begin{figure}[!t]
  \centering
  \begin{tabular}{@{}c@{}}
    \begin{minipage}{\linewidth}
      \centering
      \includegraphics[width=\linewidth]{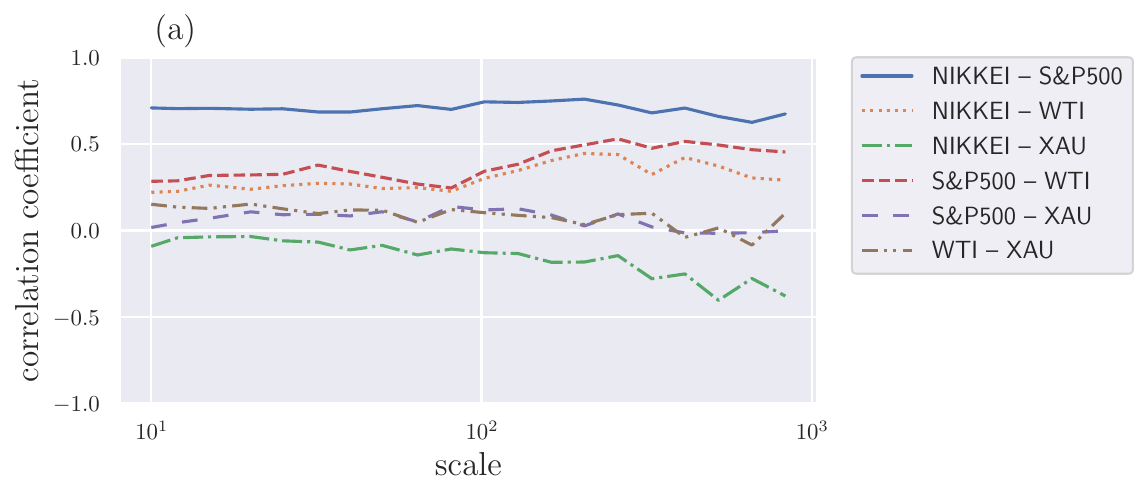}
    \end{minipage} \\
    \begin{minipage}{\linewidth}
      \centering
      \includegraphics[width=\linewidth]{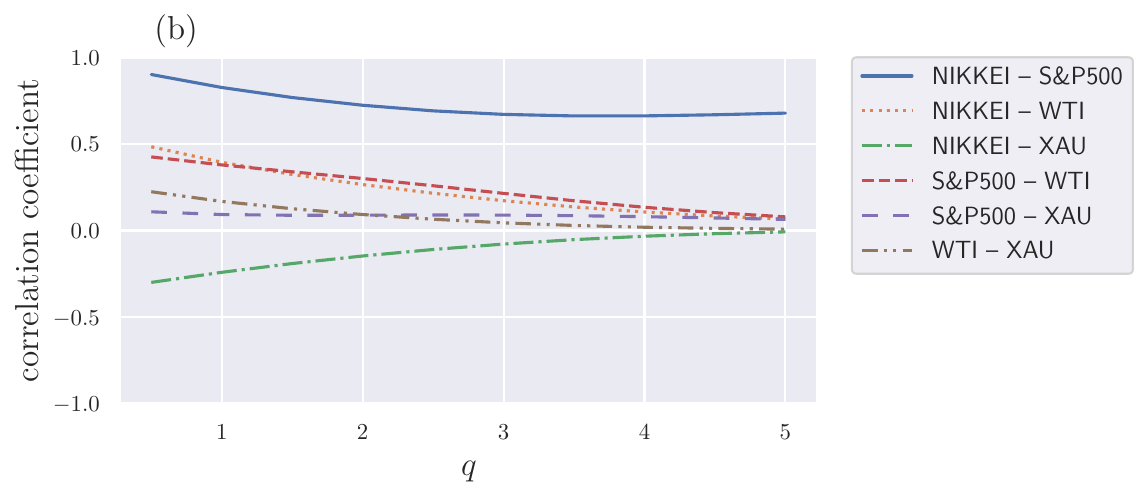}
    \end{minipage}
  \end{tabular}
  \caption{Cross-correlation properties of empirical return series. (a) The DCCA coefficient calculated across scales for each pair of daily returns over the entire period. The coefficients show scale-dependent patterns. (b) The $q$DCCA coefficient calculated across fluctuation orders $q$ at fixed scale $s=100$ for each pair of daily returns over the entire period. The coefficients vary with $q$, indicating fluctuation-order dependence in cross-correlations.}
  \label{fig:empirical_crosscorr}
\end{figure}

We also examine multiscale and amplitude-dependent characteristics of cross-correlations between assets.
If there is no multiscale property in the cross-asset relationship, the cross-correlation coefficient should not show systematic variation across scales. In Figure~\ref{fig:empirical_crosscorr}(a), the DCCA coefficients~\cite{zebende2011dcca} tend to vary with $s$, implying that a heterogeneously diversified effect can be expected when constructing a portfolio.
Similarly, by examining the cross-correlation coefficients for different fluctuation orders $q$, we examine whether the cross-asset dependence varies with the fluctuation amplitude. As can be seen in Figure~\ref{fig:empirical_crosscorr}(b), the $q$DCCA coefficient~\cite{Kwapie2015} changes with $q$ at the fixed scale $s=100$, so small and large fluctuations are cross-correlated to different degrees. We read this $q$-dependence as amplitude dependence of the cross-correlations. The coefficient itself is computed at a single scale and does not rely on any power-law behavior of the fluctuation functions, so it can be applied to nonfractal signals as well~\cite{Kwapie2015, Watorek2025}. Multifractal cross-correlations would in addition require a $q$-dependent exponent, which the portfolio criterion does not presuppose.
These results suggest that incorporating heterogeneous scales and hierarchical fluctuation orders can provide additional information for effective portfolio diversification. This further motivates us to investigate the performance of the MMFC portfolio with real-world financial data.

\begin{figure*}[!t]
  \centering
  \begin{tabular}{@{}c@{}}
    \begin{minipage}{0.65\linewidth}
      \centering
      \includegraphics[width=\linewidth]{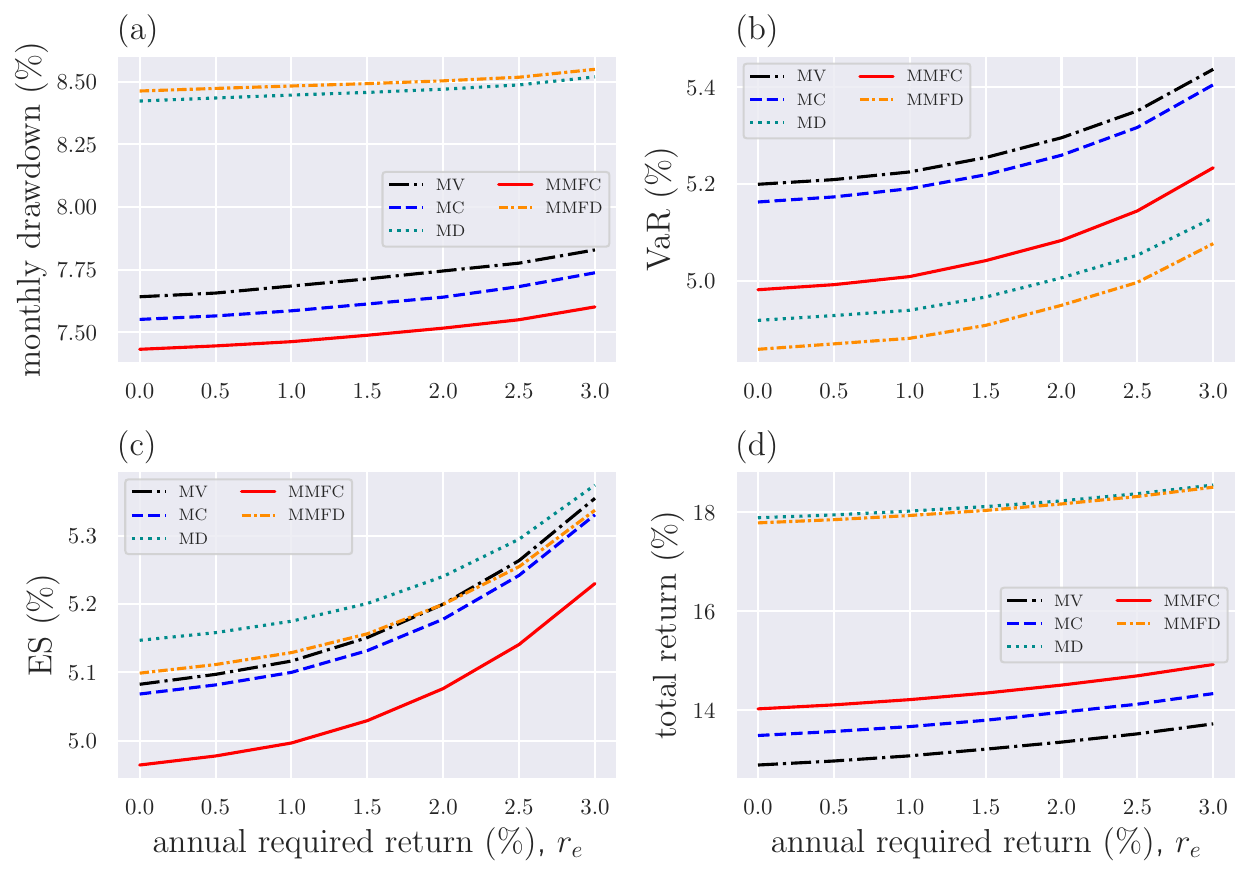}
    \end{minipage}
  \end{tabular}
  \caption{In-sample portfolio performance under annual required returns between 0\% and 3\%. The panels show rolling-window averages of (a) monthly drawdown, (b) 10-day 99\% VaR, (c) 10-day 97.5\% ES, and (d) total return, where total return is the cumulative sum of daily log returns over each estimation window. VaR and ES are reported as positive loss measures and are computed by the historical method from daily portfolio returns. MV denotes the conventional mean-variance portfolio, MC and MD denote the $q=2$ multiscale benchmarks of the sign-preserving MFCCA type and the absolute-value MFDCCA type, respectively, and MMFC and MMFD denote the corresponding portfolios aggregated over the selected fluctuation orders; MMFC is the proposed model.}
  \label{fig:in-sample_performance}
\end{figure*}

\begin{figure*}[!t]
  \centering
  \begin{tabular}{@{}c@{}}
    \begin{minipage}{0.65\linewidth}
      \centering
      \includegraphics[width=\linewidth]{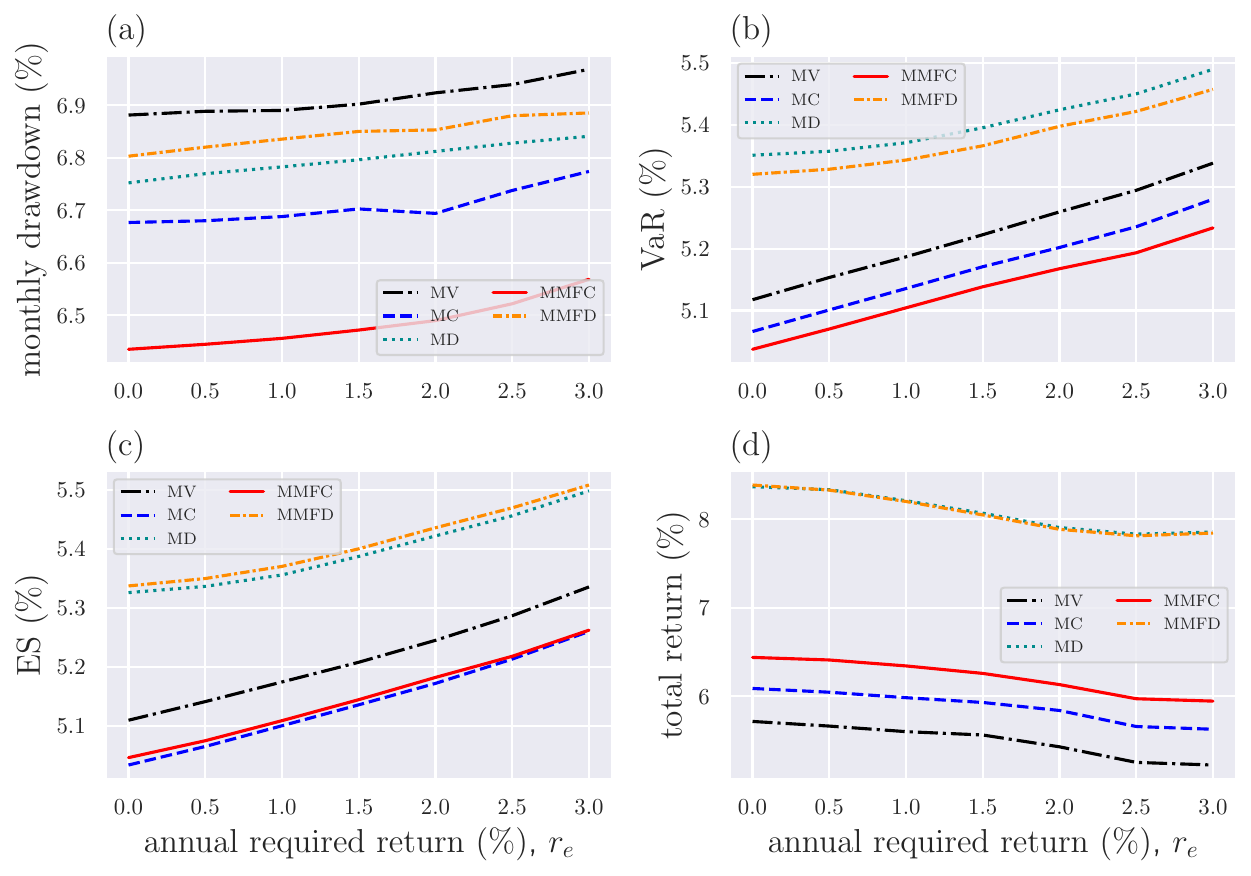}
    \end{minipage}
  \end{tabular}
  \caption{Out-of-sample portfolio performance under annual required returns between 0\% and 3\%. Portfolio weights are estimated with a rolling window of $N_w=520$ observations and rebalanced monthly. At the end of each estimation window, portfolio weights are computed and then held fixed over the following month. The window is then rolled forward by one month.
  The panels show realized out-of-sample (a) monthly drawdown, (b) 10-day 99\% VaR, (c) 10-day 97.5\% ES, and (d) total return, where total return is the annualized average of yearly cumulative log returns. VaR and ES are reported as positive loss measures and are computed by the historical method from daily portfolio returns. MV denotes the conventional mean-variance portfolio, MC and MD denote the $q=2$ multiscale benchmarks of the sign-preserving MFCCA type and the absolute-value MFDCCA type, respectively, and MMFC and MMFD denote the corresponding portfolios aggregated over the selected fluctuation orders; MMFC is the proposed model.}
  \label{fig:out-of-sample_performance}
\end{figure*}

\subsection{In-sample and out-of-sample portfolio performance}
Given that the selected assets exhibit multiscale and amplitude-dependent cross-correlation patterns, the proposed MMFC portfolio is expected to be appropriate. Recent mean-MFDCCA-based multi-asset portfolio optimization also embeds multifractal dependence into empirical portfolio construction~\citep{Li2024MultiAssetMD}; by contrast, the present empirical comparison uses MFCCA to preserve signed detrended covariance information. To find empirical evidence of an improvement in the diversification effect, both the in-sample and out-of-sample portfolio performance are examined.
The empirical analysis is intended as a methodological demonstration of how the proposed MFCCA-based allocation can be applied to real financial returns, rather than as a fully specified trading strategy. Accordingly, the comparison focuses on whether multiscale and multifractal dependence information is reflected in realized risk measures under a common rolling-window design.

In the in-sample performance testing, we use a rolling estimation window consisting of $N_w=520$ data points (approximately 2 years). The first window ends in January 2013, and the window is then shifted forward by one month, yielding 131 windows through November 2023. For each window, the MV, MC, MD, MMFC, and MMFD portfolios are constructed to calculate the 10-day 99\% VaR and 10-day 97.5\% ES, in addition to the monthly drawdown. Expected returns $E(r_i)$ are estimated by the sample means of daily returns within each estimation window, and all constrained optimization problems are solved numerically as quadratic programs. The 10-day VaR and ES are computed by the historical method from 10-day cumulative portfolio returns constructed from daily returns, while the monthly drawdown is the largest peak-to-trough decline of the cumulative portfolio return path within each month. Following the general guideline of appropriate scale ranges for detecting scaling properties in empirical analysis~\cite{Kantelhardt2002}, the scale subset is set as $S=\{5, 30, 55, 80, 105, 130\}$, with the maximum value corresponding to the standard upper limit of $N_w/4$~\footnote{
  The smallest scale $s=5$ lies below the range recommended in Section~II for estimating scaling exponents. Here, however, the fluctuation function is used as a risk functional evaluated at fixed scales rather than for exponent estimation. We also verified that excluding $s=5$ from $S$ leaves all empirical results essentially unchanged, with the same ranking of the portfolios across all values of $r_e$.
}.

It is noteworthy that, for financial returns, focusing on relatively large magnitudes of fluctuations is favorable to identify cross-dependence in the MFCCA framework, especially in the power-law form~\cite{Owiecimka2014}. Therefore, we set the fluctuation subset to $Q=\{ 1, 1.5,2, 2.5, 3, 3.5, 4 \}$ and employ this range throughout the empirical analysis~\footnote{
  We also checked the definiteness of the estimated MFCCA fluctuation matrices over all rolling windows, scales, and fluctuation orders. Non-PSD cases accounted for only 0.95\% of all MMFC fluctuation matrices, and every non-PSD case occurred at $q=1$. These exceptional cases were projected to a PSD matrix before optimization, as described in Section II.
}.
Under different values of annual required return $r_e$ between 0\% and 3\%, the five portfolios are evaluated in Figure~\ref{fig:in-sample_performance}. The annual required return $r_e$ is converted to the corresponding daily target return before solving each optimization problem; over this range, the return constraint is feasible in every estimation window used in the analysis. For each specific $r_e$, the computed risk measures VaR and ES are averaged over all 131 in-sample windows.

The in-sample comparison reveals a clear division of strengths between the two families of fractal portfolios. The proposed MMFC portfolio, depicted by solid lines, attains the smallest average monthly drawdown and ES among all five portfolios for every $r_e$, improving upon both the MV benchmark and the $q=2$ benchmark MC. The MFDCCA-based portfolios attain slightly smaller in-sample VaR and higher in-sample returns, but at the cost of visibly larger drawdowns.
Because consecutive estimation windows overlap, we assess these differences with paired tests on the per-window measures using heteroskedasticity- and autocorrelation-consistent (HAC) standard errors. The drawdown improvement of MMFC is statistically significant against MV and against the MFDCCA-based portfolios ($p\simeq0.04$ at $r_e=1\%$), and its ES improvement over MC is also significant ($p\simeq0.01$), whereas the remaining pairwise differences are not significant at the 5\% level.
This comparison suggests that incorporating fluctuation-order dependence in addition to scale dependence strengthens drawdown and tail-risk control, while the sign-preserving construction distinguishes the MMFC portfolio most clearly from its MFDCCA-based counterparts.

\begin{table*}[!t]
  \caption{\label{tab:robustness}%
  Sensitivity of the MMFC portfolio to the inputs of Eq.~\ref{eq:optimal_weight} at $r_e=1\%$. Each row changes one input relative to the baseline $S_0=\{5,30,\ldots,130\}$, $Q_0=\{1,1.5,\ldots,4\}$ with uniform $\alpha(s)$ and $\beta(q)$; the tilted profiles are defined in the text. In-sample values are averages over the 131 estimation windows, and out-of-sample values are realized over the full backtest period. Drawdown (DD), VaR, and ES are reported as positive loss measures in percent, and total return (TR) is the out-of-sample annualized return in percent. The MV, MC, MD, and MMFD rows give the baseline benchmarks of Figs.~\ref{fig:in-sample_performance} and~\ref{fig:out-of-sample_performance}.
    }
  \begin{ruledtabular}
    \begin{tabular}{lccccccc}
      & \multicolumn{3}{c}{In-sample} & \multicolumn{4}{c}{Out-of-sample} \\
      Specification & DD & VaR & ES & DD & VaR & ES & TR \\ \hline
      MV & 7.68 & 5.22 & 5.12 & 6.89 & 5.19 & 5.17 & 5.60 \\
      MC & 7.59 & 5.19 & 5.10 & 6.69 & 5.14 & 5.10 & 5.98 \\
      MD & 8.45 & 4.94 & 5.17 & 6.78 & 5.37 & 5.36 & 8.21 \\
      MMFD & 8.48 & 4.88 & 5.13 & 6.84 & 5.34 & 5.37 & 8.20 \\ \hline
      \multicolumn{8}{l}{MMFC} \\
      \quad baseline: $S_0$, $Q_0$, uniform & 7.46 & 5.01 & 5.00 & 6.46 & 5.10 & 5.11 & 6.34 \\
      \quad $Q=\{0.5,1,\ldots,5\}$ & 7.52 & 5.01 & 5.00 & 6.38 & 5.11 & 5.14 & 6.48 \\
      \quad $Q=\{2,2.5,\ldots,4\}$ & 7.57 & 4.89 & 4.89 & 6.62 & 5.10 & 5.14 & 6.64 \\
      \quad $Q=\{1,1.5,2\}$ & 8.27 & 5.57 & 5.38 & 6.80 & 5.20 & 5.12 & 5.72 \\
      \quad $S=\{30,55,\ldots,130\}$ & 7.45 & 5.03 & 5.01 & 6.39 & 5.07 & 5.08 & 6.49 \\
      \quad $S=\{5,10,20,40,80,130\}$ & 7.59 & 5.06 & 5.01 & 6.44 & 5.11 & 5.11 & 6.22 \\
      \quad $S=\{30,80,130\}$ & 7.48 & 5.04 & 5.01 & 6.33 & 5.03 & 5.05 & 6.41 \\
      \quad $\alpha(s)$ tilted to short scales & 7.55 & 5.03 & 5.00 & 6.56 & 5.14 & 5.14 & 6.16 \\
      \quad $\alpha(s)$ tilted to long scales & 7.43 & 5.02 & 5.01 & 6.37 & 5.09 & 5.08 & 6.52 \\
      \quad $\beta(q)$ tilted to small $q$ & 7.73 & 5.19 & 5.11 & 6.54 & 5.14 & 5.10 & 6.11 \\
      \quad $\beta(q)$ tilted to large $q$ & 7.50 & 4.92 & 4.92 & 6.53 & 5.08 & 5.13 & 6.58 \\
    \end{tabular}
  \end{ruledtabular}
\end{table*}

The out-of-sample performance of the multifractal portfolio is also evaluated. As in in-sample testing, the same rolling estimation procedure is conducted. Thus, the portfolio statistics during the test period are computed using the allocations during the training period, with allocations rebalanced monthly.
Figure~\ref{fig:out-of-sample_performance} shows that, in the out-of-sample period, the MMFC portfolio attains the smallest average monthly drawdown and VaR among all five portfolios for every $r_e$, with ES essentially on par with the MC benchmark, and it improves total returns relative to MV and MC. The MFDCCA-based portfolios display the opposite profile: they deliver the highest out-of-sample returns but also the largest VaR and ES among the fractal portfolios.
We again assess the differences by paired tests. The drawdown advantage of MMFC over the MFDCCA-based portfolios is unambiguous (monthly paired tests, $p<10^{-4}$ at $r_e=1\%$), which directly reflects the value of preserving the sign of local co-movements. In contrast, the average improvements of MMFC over the MV benchmark, while consistent in sign across all $r_e$, remain within the resolution of the out-of-sample period (block-bootstrap tests for VaR and ES, paired monthly tests for drawdown), and the return advantage of the MFDCCA-based portfolios is at most marginal.
Taken together, the out-of-sample evidence indicates a risk--return trade-off between the two multifractal constructions: the sign-preserving MMFC criterion favors downside-risk control, most visibly in drawdowns, whereas absolute-value-based criteria favor realized returns. These results suggest that the signed dependence structure provides allocation-relevant information beyond the conventional covariance, although the sample may be too short to establish uniform statistical dominance over the mean--variance benchmark.

\subsection{\label{sec:robustness}Robustness to the choice of scales, fluctuation orders, and preference weights}
The aggregation in Eq.~\ref{eq:optimal_weight} depends on the scale set $S$, the fluctuation order set $Q$, and the preference profiles $\alpha(s)$ and $\beta(q)$. To see how much the empirical results rest on these choices, we change one of them at a time from the baseline of the previous subsection and repeat the in-sample and out-of-sample evaluation with the rolling-window design unchanged. Table~\ref{tab:robustness} lists the results for the MMFC portfolio (for simplicity, all portfolios are studied at $r_e=1\%$).
Three alternatives are considered for $Q$. The first is the wider set $\{0.5,1,\ldots,5\}$ as used in Sec.~\ref{subsec:multifractal}, the second is the large-fluctuation subset $\{2,2.5,\ldots,4\}$, and the third is the small-fluctuation subset $\{1,1.5,2\}$. For the alternative choices of $S$ we drop $s=5$, use a denser set $\{5,10,20,40,80,130\}$, and a sparse set $\{30,80,130\}$. For the preference profiles we follow Ref.~\cite{Kakinaka2023} and tilt the weights linearly in the parameter. A long-scale tilt sets $\alpha(s)$ proportional to $s$, so that the weight of $s_\mathrm{max}=130$ is 26 times that of $s_\mathrm{min}=5$, and a short-scale tilt assigns the same weights in reverse order. The tilts of $\beta(q)$ are defined in the same way with $q$ in place of $s$, which gives a ratio of 4 between the weights of $q_\mathrm{max}=4$ and $q_\mathrm{min}=1$.

Nine of the ten alternatives leave the ranking of the portfolios unchanged. Under every scale set, every preference profile, the wider and large-fluctuation sets of $Q$, the MMFC portfolio keeps the smallest in-sample drawdown and ES, and the smallest out-of-sample drawdown and VaR among other methods, in addition to its out-of-sample total return staying above that of MV.
One exception is when $\beta(q)$ is tilted toward small $q$, the in-sample drawdown of MMFC is $7.73\%$, slightly above the $7.68\%$ of MV.
A notable difference arises when the small-fluctuation set $Q=\{1,1.5,2\}$ is used. In this case, the in-sample drawdown and VaR of MMFC are larger than those of MV, fails to provide advantage over the conventional mean-variance benchmark.
The orders above $q=2$, which emphasize the large fluctuations, are therefore the ones that produce the gain. This agrees with the finding in \cite{Watorek2025} that the correlation structure of large fluctuations ($q=4$) differs from that of medium-sized ones ($q=2$), and with the observation that the cross-dependence of financial returns is carried by large fluctuations~\cite{Owiecimka2014}.
As long as $Q$ covers the large-fluctuation range, the form of Eq.~\ref{eq:optimal_weight} does not drive the empirical results.

\section{Conclusion}
This study proposed a mean-MFCCA portfolio model that incorporates multiscale and multifractal dependence into the mean--variance allocation framework. Instead of relying solely on the conventional variance--covariance matrix, the proposed model employs the MFCCA fluctuation function as a signed multifractal risk criterion. This formulation allows portfolio allocation to depend on both the time scale $s$ at which cross-asset dependence is measured and the fluctuation order $q$ that distinguishes small and large fluctuations, while preserving the sign of local detrended covariance in the risk evaluation.

The numerical experiments demonstrate how multiscale and multifractal dependence structures are translated into portfolio allocation. Synthetic data generated from the two-component ARFIMA model show that scale-dependent cross-correlations modify the efficient frontier, whereas Markov-switching multifractal series indicate that fluctuation-order dependence also influences allocation outcomes. When both properties are present, the proposed model achieves lower tail-risk measures than the conventional mean--variance model and the fractal-based benchmarks; paired tests across realizations confirm all pairwise ES differences. In particular, the comparison with the mean-MFDCCA model highlights the importance of preserving signed local covariance information: even the $q=2$ sign-preserving benchmark outperforms the fully multifractal absolute-value-based portfolio.

Empirical application further illustrates the usefulness of the proposed framework. Using four representative financial assets, the rolling-window analysis shows that the proposed portfolio attains the lowest average drawdown among the portfolios considered, in and out of sample, together with the lowest out-of-sample VaR, while maintaining portoflio returns. The gains are unambiguous in negative tail risk control relative to the MFDCCA-based benchmarks. These results suggest that signed multifractal dependence provides information beyond that contained in the conventional covariance risk criterion, with its practical value concentrated in the control of large adverse co-movements.

The results should nevertheless be interpreted within the scope of the present study. The empirical analysis is limited to four representative financial assets and a specific sample period, and the out-of-sample sample size limits the statistical power of the comparisons against the mean--variance benchmark. Moreover, for $q\neq2$ the proposed criterion is an aggregation of signed pairwise dependence rather than the portfolio's own multifractal fluctuation measure, and the model assumes no-short-sale constraints, fixed sets of scales and fluctuation orders, equal preference weights across $s$ and $q$, and does not explicitly account for transaction costs or liquidity constraints.
Also, the averaging of optimal weights over scales and fluctuation orders is somewhat heuristic. Alternative sets with different choices of scale, preference weights, and fluctuation orders leave the empirical findings unchanged as long as the fluctuation orders above $q=2$ are included, but how these inputs should be chosen in general is left open.

Future work may extend the framework in several directions from both physics and finance. For example, asymmetric multifractal dependence could be incorporated using MFADCCA~\citep{Cao2014MFADCCA}, while investor preferences over time scales and fluctuation orders could be allowed to evolve dynamically with market conditions.
More broadly, the proposed framework provides a mapping from signed multiscale interaction patterns to resource-allocation outcomes through multifractal cross-correlation functions.
Whether the framework also applies to other complex systems (e.g., in economics, biology, or engineering) with heterogeneous scales and amplitude-dependent coupling has to be examined separately for each system.

\begin{acknowledgments}
This work was supported by JSPS KAKENHI Grant Number JP26K24304.
\end{acknowledgments}

\section*{Data availability}
The raw one-minute quote data are publicly available from \url{http://www.histdata.com/}. The processed datasets and the analysis code that reproduce all results, including the fixed random seeds used in the simulations, are available from the corresponding author upon reasonable request.

\section*{AI use disclosure}
During the preparation of this research, the authors used Claude Code (Opus 5.5) to assist with language polishing, literature organization, and code debugging. The authors reviewed, edited, and verified all outputs and take full responsibility for the published content.

\bibliography{references}

\end{document}